\documentclass{article}
\usepackage{amsmath,graphicx,mlspconf}
\usepackage{amsmath, graphicx}
\usepackage{amssymb}
\usepackage{booktabs, tabularx}
\usepackage{algorithm}
\usepackage{algpseudocode}
\usepackage{siunitx}

\copyrightnotice{U.S.\ Government work not protected by U.S.\ copyright}

\copyrightnotice{979-8-3503-2411-2/23/\$31.00 {\copyright}2023 Crown}

\copyrightnotice{979-8-3503-2411-2/23/\$31.00 {\copyright}2023 European Union}

\copyrightnotice{979-8-3503-2411-2/23/\$31.00 {\copyright}2023 IEEE}

\toappear{2023 IEEE International Workshop on Machine Learning for Signal Processing, Sept.\ 17--20, 2023, Rome, Italy}

\title{IANS: Intelligibility-aware Null-steering Beamforming for Dual-Microphone Arrays}
\name{Wen-Yuan Ting$^{\star}$, Syu-Siang Wang$^{\dagger}$, Yu Tsao$^{\ddagger}$, and Borching Su$^{\star}$}
\address{$^{\star}$ Graduate Institute of Communication Engineering, National Taiwan University, Taipei, Taiwan \\$^{\dagger}$ Department of Electrical Engineering, Yuan Ze University, Taoyuan, Taiwan \\
$^{\ddagger}$ Research Center for Information Technology Innovation, Academia Sinica, Taipei, Taiwan}

\begin{document}
\ninept

\maketitle

\begin{abstract}
Beamforming techniques are popular in speech-related applications due to their effective spatial filtering capabilities. Nonetheless, conventional beamforming techniques generally depend heavily on either the target's direction-of-arrival (DOA), relative transfer function (RTF) or covariance matrix. This paper presents a new approach, the intelligibility-aware null-steering (IANS)  beamforming framework, which uses the STOI-Net intelligibility prediction model to improve speech intelligibility without prior knowledge of the speech signal parameters mentioned earlier. The IANS framework combines a null-steering beamformer (NSBF) to generate a set of beamformed outputs, and STOI-Net, to determine the optimal result. Experimental results indicate that IANS can produce intelligibility-enhanced signals using a small dual-microphone array. The results are comparable to those obtained by null-steering beamformers with given knowledge of DOAs.
\end{abstract}
\section{Introduction}
\label{sec:intro}


Microphone arrays are commonly used in numerous speech-related applications including hearing aids and teleconferencing to isolate the desired signals that are often degraded
by ambient noise and other types of interference \cite{chakraborty2013joint,kumatani12_interspeech}.
Multi-channel speech enhancement (MCSE) techniques have been extensively studied to extract the desired signals
\cite{ochiai2017unified}.
Beamforming algorithms are usually a crucial component of these methods, as they utilize spatial diversity from multiple recordings to perform spectral and spatial filtering on multiple channel inputs, generating a speech-enhanced output
\cite{cox1987robust,wang2022study}.
For example, the delay-and-sum beamformer 
\cite[Chapter 3]{bai2013acoustic} uses the geometry of the array and direction-of-arrival (DOA) information to parameterize the spatial-spectral filter. 
The minimum variance distortionless response (MVDR) method \cite{capon1969high} 
minimizes the power of the noise signal while maintaining a distortionless response for the target signal, utilizing the knowledge of the covariance matrices and DOA or relative transfer function (RTF).
Additionally, null-steering beamformers (NSBF) have been proposed to 
filter out signals from specific directions
\cite{schepker2016acoustic, ikram2002beamforming,li2018beamforming}.


Conventional beamforming algorithms typically depend highly on an accurate DOA or RTF estimate to obtain the spatial information of the target signals. 
Over the past few decades, multiple DOA estimation algorithms have been proposed in 
\cite{schmidt1986multiple,roy1989esprit, tho2014robust}.
For DOA estimation algorithms specialized for multiple speech signals, the work in \cite{tho2014robust} used the coherence test and sparsity property of speech 
to estimate accurate DOAs using clustering-based methods.
In addition to a direct DOA estimation approach
, time difference of arrival (TDOA) estimation methods \cite{dibiase2000high,omologo1997use,knapp1976generalized} are also commonly used to localize the target signal. 
One popular category is the application of the steered response power phase transform \cite{dibiase2000high}, which scans over a predefined spatial region to parameterize the cross-correlation functions using each candidate location of the source, and then adopts a maximum likelihood estimator to estimate the TDOA.
In addition to these methods, the work in \cite{markovich2015performance} discussed covariance subtraction and covariance whitening methods to obtain RTF estimates of the speech signal using well-estimated covariance matrices from noise-only and speech-noise frames. 
Although these approaches have great potential to provide accurate spatial information, they typically rely heavily on multiple assumptions. In the case of \cite{markovich2015performance}, the authors assumed accurate estimates of the noise covariance matrices for each time-frequency index. 
If the noise covariance matrices 
contain spatial statistics of the speech signal, the beamformers might not be aware of such errors and attenuate the corresponding signals without regard to how this might impact the intelligibility of speech signals.
Meanwhile, it is also worth noting that, 
neural beamformers, such as \cite{ochiai2017unified, drude2018integrating, heymann2016neural, pfeifenberger2019deep}, have been proposed to perform state-of-the-art MCSE.
For these NN-based approaches, it is usually necessary to construct a dataset containing diverse utterances received by a microphone array in multiple scenarios. 
In addition, 
these neural beamformers are usually optimized over a large number of parameters, which makes each parameter hard to interpret. 

In the field of speech processing, 
a well-known metric for intelligibility is the short-time objective intelligibility (STOI) \cite{taal2011algorithm}. 
The STOI function estimates the intelligibility of signals through 
a series of signal-processing stages, including silence-segment elimination, feature extraction in the time-frequency (TF) domain, one-third octave band processing, feature normalization, and intelligibility mapping. In this process, the deteriorated sound signal and the corresponding clean reference signal are used simultaneously to compute the final score. 
In this paper, the STOI function will be denoted as $\boldsymbol{\mathit{h}}_\mathrm{STOI}: \mathbb{R}^{N \times K} \times \mathbb{R}^{N \times K} \rightarrow [0, 1]$ which is defined as the mapping from the magnitude of a pair of $N \times K$ short-time Fourier transform (STFT) matrices to the interval $[0, 1]$, where $N$ is the number of time frames and $K$ is the number of frequency bins per frame. For simplicity, we will omit the steps such as silence-segment elimination for our description of the STOI function.
In addition, unlike metrics such as the speech intelligibility index in \cite{ansi1997methods}, STOI is known for its reliable intelligibility evaluation of signals processed in the TF domain, where most acoustic beamforming systems perform MCSE. However, a clean reference is typically inaccessible. Therefore,
the authors in \cite{zezario2020stoi} proposed STOI-Net, a non-intrusive intelligibility assessment model that predicts STOI scores based only on the noise-corrupted waveforms.

In light of the heavy dependence of beamformers on the estimation of the DOAs or RTFs of the speech signals, 
we propose a new optimization framework for an intelligibility-enhancing beamformer 
without relying on the previously mentioned speech parameters. 
Instead, we explicitly consider intelligibility as an optimization objective. 
Works such as \cite{ward1999beamforming} have also incorporated the notion of intelligibility into the design of beamformers. We will perform intelligibility-based optimization within a set of null-steering beamformers. Hence, we call this intelligibility-aware null-steering (IANS) beamforming.
For the IANS beamforming process, an NSBF algorithm is first applied to generate a set of candidate signals via null-steering. The generated signals are then passed through a pre-trained STOI-Net to predict the associated STOI scores. IANS then outputs the utterance corresponding to the highest intelligibility score. Contrary to the previously mentioned neural beamformers, the proposed IANS algorithm doesn't require additional multi-channel training data.
Moreover, the IANS optimization problem only optimizes one parameter whose optimal value is interpretable. Furthermore, advanced single-channel SE methods, such as \cite{luo2019conv,liu2019divide, le2021dpcrn, fu2021metricgan+, lu2022espnet}, can be incorporated with IANS for downstream applications.

The remainder of this paper is organized as follows. Section \ref{sec:related_works} discusses the signal model and related works including filter-and-sum beamformers, null-steering beamformers and STOI-Net.
Next, we will present our IANS optimization problem in Section \ref{sec:prob_form}.
In Section \ref{sec:IANS_algo}, the IANS algorithm will be discussed in detail.
Section \ref{sec:exp_results} presents the experimental setup and results. Finally, Section \ref{sec:conclusion} concludes the paper and discusses future works.


\vspace{-0.3cm}
\section{Background and Related Works}
\label{sec:related_works}
\vspace{-1mm}
\subsection{Signal model}
In this study, the considered signal model comprises a 
speech signal, $s(t)$, and an interference signal, $i(t)$, propagating in a room with a sound speed of $c$, received by a dual-microphone array at angles of $\theta_s$ and $\theta_i$, respectively. The angles are
measured with respect to the first (reference) microphone, with $0^\circ$ being the endfire direction. The microphone array has a small spacing of $\ell$, and we assume that the sound sources are stationary in space. We denote the room impulse responses (RIRs) for $s(t)$ and $i(t)$ with respect to the $m^\text{th}$ microphone as
$g^{(m)}_s(t)$ and $g^{(m)}_i(t)$, respectively. The received signal at the $m^\text{th}$ microphone can be expressed as the following:

\begin{equation}\label{eq:sig_model}
    x^{(m)}(t) = g^{(m)}_s(t) \ast s(t) + g^{(m)}_i(t) \ast i(t).
\end{equation}
After obtaining the received signals $x^{(1)}(t)$ and  $x^{(2)}(t)$, We can then apply the STFT to derive their corresponding $N \times K$ STFT matrices $\mathbf{X}^{(1)}$ and $\mathbf{X}^{(2)}$. Subsequently, we can define the received signal vector $\mathbf{x}[n, k]$ as follows:

\begin{equation}
    \mathbf{x}[n, k] = [\mathbf{X}^{(1)}_{n, k}, \mathbf{X}^{(2)}_{n, k}]^T.
\end{equation}
Here, $\mathbf{X}^{(m)}_{n, k}$ represents the $(n, k)^{\text{th}}$ element of $\mathbf{X}^{(m)}$, where
$n=1, 2,  \cdots, N$ and $k=1, 2,  \cdots, K$.


\vspace{-1mm}
\subsection{Filter-and-sum beamformers}
Filter-and-sum beamformers \cite{frost1972algorithm} 
are a set of beamformers that perform the filter-and-sum operation to enhance the signal of interest. This process can be represented in the TF-domain as

\begin{equation}
    Y[n, k] = \mathbf{w}^H[n, k] \mathbf{x}[n, k],
\end{equation}
where $\mathbf{w}[n, k]$ is the weight vector for $\mathbf{x}[n,\: k]$, and $Y[n, k]$ is the resulting TF component of the beamformed signal. We will denote this set of beamformers as $\boldsymbol{\mathcal{F}}_\mathrm{FSBF}$. 
\vspace{-2mm}
\subsection{Null-steering beamformers}
Within $\boldsymbol{\mathcal{F}}_\mathrm{FSBF}$, there is a subset of beamformers capable of nulling out signals coming from a particular direction $\phi$ while maintaining a (nearly) distortionless response at $\theta_d$. We call this set the null-steering beamformer set, $\boldsymbol{\mathcal{F}}_\mathrm{NSBF}$. 

We first define two vectors, the distortionless response steering vector $\mathbf{a}^{(\theta_d)}[k] = [1, e^{-j \frac{\omega_k \ell}{c} \cos{\theta_d}}]^{T}$ and the null-response steering vector
$\mathbf{a}^{(\phi)}[k] = [1, e^{-j \frac{\omega_k \ell}{c} \cos{\phi}}]^{T}$, where $\omega_k$ is the frequency value at the $k^\text{th}$ frequency bin. Each $\mathbf{a}^{(\phi)}[k]$ is associated with a projection matrix $\mathbf{\Phi}[k]$ defined in the following,

\begin{equation}\label{eq:Phi}
    \begin{aligned}
         \mathbf{\Phi}[k]&=\mathbf{I} - \frac{\mathbf{a}^{(\phi)}[k](\mathbf{a}^{(\phi)}[k])^H}{||\mathbf{a}^{(\phi)}[k]||^2} \\
         &=\mathbf{I} - \frac{\mathbf{a}^{(\phi)}[k](\mathbf{a}^{(\phi)}[k])^H}{2},
    \end{aligned}
\end{equation}
where $(\cdot)^H$ denotes the Hermitian transpose operator. Here, $\mathbf{\Phi}[k]$ projects vectors into the subspace orthogonal to the span of $\mathbf{a}^{(\phi)}[k]$. 
In this paper, $\boldsymbol{\mathcal{F}}_\mathrm{NSBF}$ is defined as a beamformer set
with time-invariant weight vectors defined as 

\begin{equation}\label{eq:NSBF_w}
    \mathbf{w}[k] = \frac{\mathbf{\Phi}[k]\mathbf{a}^{(\theta_d)}[k]}{\max((\mathbf{a}^{(\theta_d)}[k])^H\mathbf{\Phi}[k]\mathbf{a}^{(\theta_d)}[k], \epsilon)},
\end{equation}
where $\epsilon$ is a small number to avoid 0 division. We note that without the $\max(\cdot , \epsilon)$, Eq. \eqref{eq:NSBF_w} has been studied in \cite{habets2009new} in the context of the MVDR beamformer.
It is also worth pointing out that in the context of beamformers such as the linearly-constrained minimum variance beamformer \cite{frost1972algorithm}, null responses are usually set as explicit constraints to a noise power minimization problem. 

\vspace{-2mm}
\subsection{STOI-Net}\label{subsec:STOI-Net}
\vspace{-1.5mm}
In \cite{zezario2020stoi}, the authors proposed STOI-Net, a non-intrusive speech intelligibility assessment model that predicts the STOI scores of speech signals both frame- and utterance-wise using feature extraction and score calculation functions. 
For the feature extraction, the STFT is applied to convert the peak-normalized time-domain waveform of interest into a sequence of frame-wise magnitude spectra in the frequency domain. These frames are then passed through 12 fully convolutional neural network layers to extract the acoustic representations. Next, the score-calculation function maps the extracted features to an intelligibility score. Specifically, frame-level scores are generated after applying 1) bidirectional long short-term memory, 2) an attention layer, and 3) fully connected nonlinear mapping functions to the extracted features. The final intelligibility score of the entire utterance is then obtained by applying a global averaging algorithm to all frame-level scores. It is worth noting that STOI-Net is not limited to a single neural network architecture. In \cite{zezario2020stoi}, two model architectures were used: one with an attention layer and the other without it. In the remainder of this paper, we will denote the STOI-Net model as a function $\boldsymbol{\mathit{h}}_\mathrm{STOI-Net}: \mathbb{R}^{N \times K} \rightarrow \mathbb{R}$.
\section{Problem Formulation}\label{sec:prob_form}

Contrary to most optimal beamformers for speech enhancement (e.g., MVDR) where optimal weights are derived for each TF bin based on well-estimated DOAs, RTFs and covariance matrices, we propose an optimization problem based on the intelligibility of the entire utterance of the received speech signals. 

Our primary goal is to identify a function $\boldsymbol{\mathit{f}}: \mathbb{C}^{N \times K} \times \mathbb{C}^{N \times K} \rightarrow \mathbb{C}^{N \times K}$ within a function set $\boldsymbol{\mathcal{F}}$ that takes the received signals $\mathbf{X}^{(1)}$ and $\mathbf{X}^{(2)}$ as input 
and maps them to an STFT matrix 
with the maximum STOI score. 
As we aim to perform optimization without having to train a new NN that learns from a dataset containing microphone array recordings from various scenarios, we will use a simpler function set $\boldsymbol{\mathcal{F}}=\boldsymbol{\mathcal{F}}_\mathrm{NSBF}$
where we can limit all potential values of $\theta_d$ and $\phi$ on a discrete grid. 
However, the number of feasible solutions grows quadratically with the resolution for  the $\theta_d$ and $\phi$ axes on this grid. Hence, we only perform grid search for $\phi$ on a grid $\mathcal{G}$ while fixing $\theta_d$ to an arbitrary angle $\psi \in [0^\circ, 180^\circ]$. The grid $\mathcal{G}$ is an ordered set containing $P$ angles ranging from $0^\circ$ to $180^\circ$. 
Thus, the number of possible beamformers we are considering here is $P$.
Our optimization problem can now be described as the following:

\begin{equation}\label{eq:opt_prob_2}
    \begin{aligned}
        &\mathop{\text{maximize}}_{
        \phi \in \mathcal{G}} &&\quad \boldsymbol{\mathit{h}}_\mathrm{STOI}(|\boldsymbol{\mathit{f}}_{\theta_d=\psi, \phi}(\mathbf{X}^{(1)}, \mathbf{X}^{(2)})|, |\mathbf{S}|)\\ 
        &\text{\:subject \:to} &&\quad \boldsymbol{\mathit{f}}_{\theta_d=\psi, \phi} \in \boldsymbol{\mathcal{F}}_\mathrm{NSBF},
    \end{aligned}
\end{equation}
where $\mathbf{S}$ represents the STFT matrix of the clean speech signal and $|\cdot|$ denotes the element-wise magnitude extraction for a matrix.
We refer to this problem as the STOI Null-steering (STOI-NS) problem, as it employs null-steering to optimize the true STOI function. We will denote the optimal beamformer and null angle for this problem as $\boldsymbol{\mathit{f}}_\mathrm{STOI-NS}^\star$ and $\phi_\mathrm{STOI-NS}^\star$, respectively.

However, $\mathbf{S}$ is never accessible in practical scenarios. Therefore, using the pre-trained STOI-Net model $\boldsymbol{\mathit{h}}_\mathrm{STOI-Net}$, we modify the optimization problem in \eqref{eq:opt_prob_2} as follows:
\begin{equation}\label{eq:opt_prob_3}
    \begin{aligned}
        &\mathop{\text{maximize}}_{\phi \in \mathcal{G}} &&\quad \boldsymbol{\mathit{h}}_\mathrm{STOI-Net}(|\boldsymbol{\mathit{f}}_{\theta_d=\psi, \phi}(\mathbf{X}^{(1)}, \mathbf{X}^{(2)})|)\\ 
        &\text{\:subject \:to} &&\quad \boldsymbol{\mathit{f}}_{\theta_d=\psi, \phi} \in \boldsymbol{\mathcal{F}}_\mathrm{NSBF} 
    \end{aligned}
\end{equation}
Since STOI-Net was trained to estimate the STOI score of a signal, we consider this the Intelligibility-aware Null-steering (IANS) problem.
The IANS problem is now feasible without the clean reference $\mathbf{S}$. The optimal beamformer and null angle for this problem are denoted as 
$\boldsymbol{\mathit{f}}_\mathrm{IANS}^\star$ and $\phi_\mathrm{IANS}^\star$, respectively. 
It is clear that the STOI score of the output obtained from using the beamformer $\boldsymbol{\mathit{f}}_\mathrm{STOI-NS}^\star$ is a natural upper bound of that using $\boldsymbol{\mathit{f}}_\mathrm{IANS}^\star$ as we will show in  
Section \ref{sec:exp_results}.
This optimization framework was inspired by works such as \cite{fu2021metricgan+}, 
where the authors trained speech enhancement systems by incorporating speech quality prediction neural networks \cite{fu2018quality} into the loss function.

Notably, contrary to conventional beliefs where it is generally thought to be necessary for $\psi$ to be close to $\theta_s$ in order to perform speech enhancement, our method as we will show later in Section \ref{sec:exp_results} is no longer constrained to this requirement. Therefore, we do not regard $\psi$  as an estimate of $\theta_s$.
This also implies that, in the context of the STOI-NS and IANS optimization problem, we never guarantee the distortionless property of speech as in the MVDR beamformer. However, as we will show later in Section \ref{sec:exp_results}, intelligibility enhancement is still possible using the optimal null angles 
$\phi^\star_\mathrm{STOI-NS}$ and $\phi^\star_\mathrm{IANS}$.  
These two angles 
can be interpreted as optimal null angles chosen to minimize the 
impact of 
the interference signal, speech distortion and RIRs on intelligibility, while maintaining a nearly distortionless response at $\psi$.
Since there is a chance that  $\psi=\theta_i$, 
it is advisable to perform the IANS algorithm twice using two different $\theta_d$ values (e.g., $\psi$ and $\psi + 80^\circ$ in this study). 

It is worth noting that dual-microphone array beamformers usually correspond to beampatterns exhibiting a large main lobe and side lobe owing to the limited degrees of freedom. In other words, we can use this property to construct a directive null, as in Eq. \eqref{eq:Phi}, while preserving a certain amount of gain for signals coming from all angles, except those within the vicinity of the null.
We also note that small microphone arrays tend to have frequency-invariant beampatterns as explained in \cite{huang2020differential}, which can also be an advantage since beamformers that are sensitive to frequency variations tend to produce more unpredictable results.

\section{The IANS Algorithm}\label{sec:IANS_algo}
\begin{figure}[!t]
    \centering
    \includegraphics[width=0.9\columnwidth]{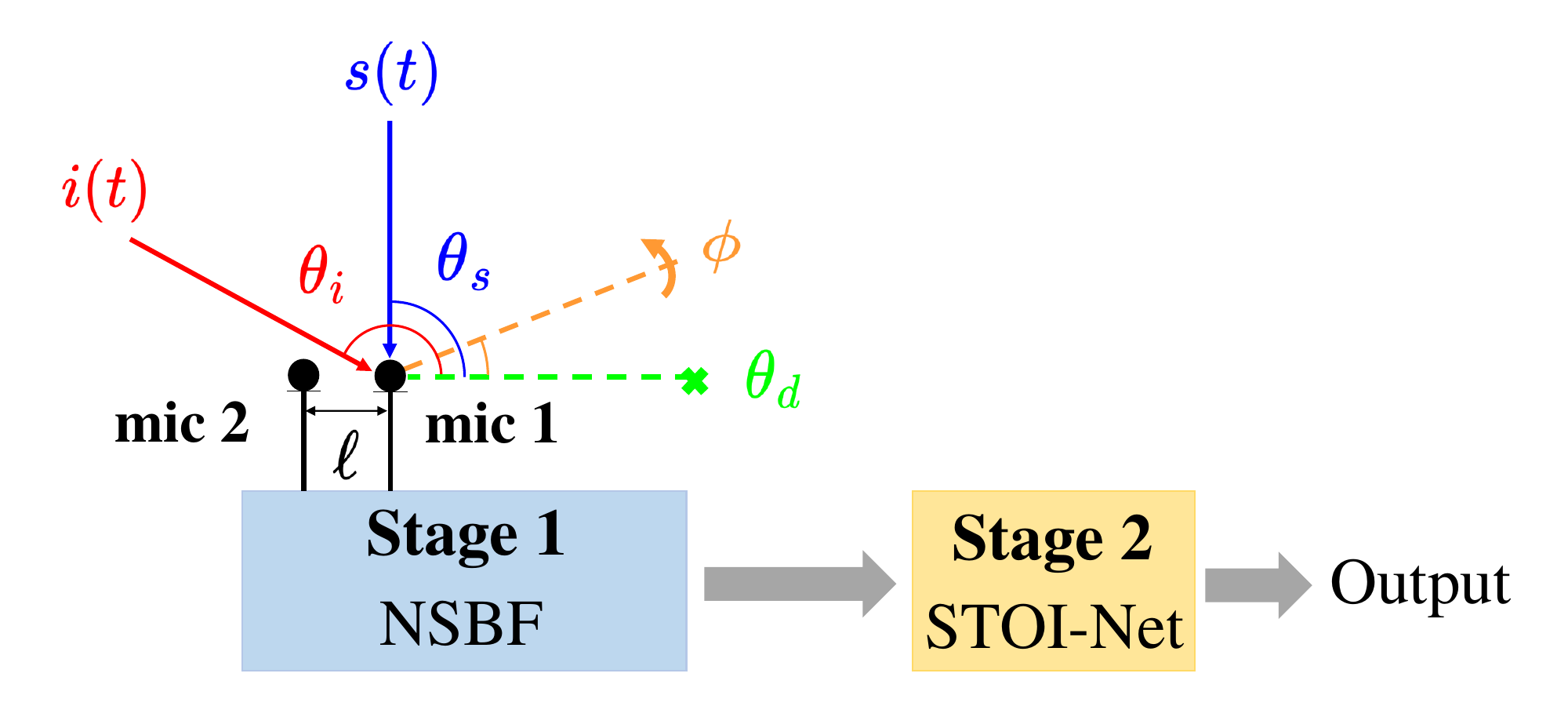}
     \vspace{-0.25cm}
    \caption{IANS block diagram}
     \vspace{-5.5mm}
    \label{fig:IANS_block_diagram}
\end{figure}

This section describes the IANS algorithm which solves the IANS optimization problem in \eqref{eq:opt_prob_3}. The algorithm consists of two stages: the NSBF stage and the STOI-Net stage as shown in Fig. \ref{fig:IANS_block_diagram},
where results from the first stage will be passed on to the second stage. The following subsections will provide more detailed explanations about the IANS algorithm.

\vspace{-2mm}
\subsection{Stage 1: NSBF}\label{subsec:NSBF}
\vspace{-1.5mm}
The initial step of the IANS algorithm involves applying the STFT on the two signals $x^{(1)}(t)$ and $x^{(2)}(t)$ to obtain $\mathbf{X}^{(1)}$ and $\mathbf{X}^{(2)}$.
We then generate a set $\mathcal{Y}_\mathrm{(STFT)}$ containing $P$ STFT matrices $\{ \mathbf{Y}^{(1)}, \cdots \mathbf{Y}^{(P)} \}$ by sending the pair $(\mathbf{X}^{(1)}, \mathbf{X}^{(2)})$ into $P$ NSBF beamformers $\{\boldsymbol{\mathit{f}}_{\theta_d =\psi, \phi=\mathcal{G}_1}, \cdots ,\boldsymbol{\mathit{f}}_{\theta_d =\psi, \phi=\mathcal{G}_{P}}\}\subset\boldsymbol{\mathcal{F}}_\mathrm{NSBF}$. If $\psi = \mathcal{G}_p$, where $p\in \{1, 2, \cdots, P\}$, we let $\mathbf{Y}^{(p)}=\mathbf{X}^{(1)}$ instead of using 
$\boldsymbol{\mathit{f}}_{\theta_d =\psi, \phi=\psi}$. 
Note that parallel computing can be used since each computation of the elements of $\mathcal{Y}_\mathrm{(STFT)}$ is independent of each other. It is also worth pointing out that the time-invariant weight vectors in Eq. \eqref{eq:NSBF_w} can be computed and stored beforehand to save time. 

Since we will later send these into STOI-Net, we apply the inverse-STFT operation (iSTFT) on each element in $\mathcal{Y}_\mathrm{(STFT)}$ to perform peak normalization in the time domain. We denote this set as $\mathcal{Y}_\mathrm{(time)}'$. We do this to match the training conditions of STOI-Net as we mentioned in Subsection \ref{subsec:STOI-Net}.

\vspace{-2mm}
\subsection{Stage 2: STOI-Net}
\vspace{-1.5mm}

Following the peak normalization, we perform STFT on each element in $\mathcal{Y}_{\mathrm{(time)}}'$ to convert them back to the TF domain and extract their corresponding magnitude components. We denote the resulting set as $\mathcal{Y}_{\mathrm{(STFT)}}''$.
We then pass each element in $\mathcal{Y}_{\mathrm{(STFT)}}''$ into STOI-Net to predict their utterance-based STOI score. These scores are then stored in a STOI-Net score vector $\boldsymbol{\alpha}$. The optimal null angle $\phi_\mathrm{IANS}^\star$ for $\boldsymbol{\mathit{f}}_\mathrm{IANS}^\star$ can be obtained as 

\begin{equation}
    \phi_\mathrm{IANS}^\star = \mathcal{G}_{\mathrm{argmax} (\boldsymbol{\alpha})}.
\end{equation}
Moreover, in the case where we have access to the clean reference signal $\mathbf{S}$, we can replace STOI-Net with the real STOI function in this stage and obtain a STOI score vector $\boldsymbol{\beta}$. Therefore, the value of 
$\phi_\mathrm{STOI-NS}^\star$ for $\boldsymbol{\mathit{f}}_\mathrm{STOI-NS}^\star$ can be expressed as 

\begin{equation}
    \phi_\mathrm{STOI-NS}^\star = \mathcal{G}_{\mathrm{argmax} (\boldsymbol{\beta})}.
\end{equation}

In this study, the pre-trained STOI-Net model without the attention layer was directly obtained from the previous research\footnote{https://github.com/dhimasryan/STOI-Net} without any modifications, such as adaptation, retraining, or fine-tuning, for the MCSE task.
\vspace{-0.2cm}
\section{Experimental results and analysis}\label{sec:exp_results}
\vspace{-0.1cm}
         
\subsection{Experimental setup}
\vspace{-1.5mm} 
In this study, the Pyroomacoustics package \cite{scheibler2018pyroomacoustics} was utilized to simulate the signal model in Eq. \eqref{eq:sig_model} using the image source method \cite{allen1979image} with the following parameters. The simulated room has dimensions of $[\SI{5}{m}, \SI{6}{m}, \SI{4}{m}]$ with the RT60 parameter set to $\SI{150}{ms}$ and the sound speed $c$ set to $\SI{343}{m/s}$. The center of the microphone array is located at $[\SI{2.5}{m}, \SI{3}{m}, \SI{1}{m}]$. The distance between the two microphones is set to $\ell=\SI{8}{mm}$ with the array being parallel to the x-axis and the reference microphone being the microphone on the right.
The speech DOA $\theta_s$ is set to $90^{\circ}$, while the interference DOA $\theta_i$ can be one of four predefined directions: $22.5^{\circ},\:67.5^{\circ},\:112.5^{\circ}$, or $157.5^{\circ}$.
IANS then uses 512-point Hamming windows with 
$50\%$ overlap to process the incoming signals.
The set $\mathcal{G}$ is a uniform
grid over the interval [$0^\circ$, $180^\circ$] with an angular resolution of $2^\circ$ (i.e., 91 angular values).
Additionally, IANS was evaluated using two values for $\psi$: $0^{\circ}$, representing the largest angle difference from $\theta_s$, and $80^{\circ}$, which is relatively closer to $\theta_s$. 
Note that the values of $\ell$ and $c$ are assumed known to the IANS algorithm. Additionally, the value of $\epsilon=1.11 \times 10^{-16}$.

Our experiments can be classified into two parts. 
The first part uses an English dataset, namely, the Wall Street Journal \cite{paul1992design} eval92 evaluation set. 
From eval92, we first selected two male and two female speakers, and chose one utterance from each speaker to form the source signal. Babble and car noises in the NOISEUS corpus \cite[Chapter 12]{loizou2007speech} and pink noise in NOISEX-92 \cite{varga1993assessment} were applied as the interference. Five signal-to-interference ratios (SIRs), namely, $-10$, $-5$, $0$, $5$, and $10$ dB, were used to create noisy utterances. The SIRs were mixed with respect to the first microphone as suggested by the Pyroomacoustics documentation. 
Therefore, 240 source interference pairs (four clean utterances, three noises, four interference angles, and five SIRs) were used to form the English testing set (denoted as ``WSJ''). 
For the second part of experiments, we used a Mandarin dataset, namely, the Taiwan Mandarin Hearing in Noise Test \cite{huang2005development} corpus, comprising 320 sentences. Two male speakers and two female speakers were selected from the dataset. One utterance, recorded in a noise-free environment, was selected from each speaker as the speech source.
For the interference signal, we chose three noise signals from the DEMAND \cite{thiemann2013diverse} dataset: ``tmetro", ``pstation", and ``npark." 
Like in WSJ, 
the aforementioned SIRs were used to create 240 source-interference pairs (four clean utterances, three noises, four interference angles, and five SIRs) for the Mandarin testing set (denoted as ``TMHINT''). It is worth noting that STOI-Net was previously trained on the training set of the original Wall Street Journal dataset. The eval92 set was used to evaluate the generalization performance of STOI-Net. 
Therefore, the English and Mandarin datasets in this study correspond to the matched and mismatched languages for STOI-Net, respectively. All single-channel recordings were sampled at $\SI{16}{kHz}$.
The experimental performance was evaluated in terms of STOI 
and the wideband extension of the perceptual evaluation of speech quality (PESQ) \cite{rix2001perceptual, miao_wang_2022_6549559} metric. 

\vspace{-2mm}
\subsection{Evaluation results}
\vspace{-1.5mm}

For both testing sets, we labeled the enhanced results using the IANS algorithm with 
$\theta_d=\psi$ as ``IANS$_{\psi}$.'' 
Noisy utterances (labeled as ``Noisy") received by the first microphone were used as the baseline. 
Moreover, we also compared our IANS result with two additional systems. The first system
is the STOI-NS system which optimizes the STOI-NS problem given the clean reference $\mathbf{S}$ for all utterances. The optimization procedure was detailed in Section \ref{sec:IANS_algo}. 
Like in ``IANS$_{\psi}$", we represent the results from STOI-NS with 
$\theta_d=\psi$ as ``STOI-NS$_{\psi}$.'' For the second system, NSBF was performed by setting $\theta_d=\theta_s$ and $\phi=\theta_i$. This system has the advantage of knowing the true DOAs of the speech and interference signals. Therefore, we label the corresponding results as ``T-NSBF."


For the WSJ evaluation set, we list the average STOI and PESQ scores for all 240 utterances of ``Noisy'', ``IANS$_{0^\circ}$'', ``STOI-NS$_{0^\circ}$'', and ``T-NSBF'' in Table \ref{tab:expI_wsj}. From the table, we can see that the STOI and PESQ scores for ``IANS$_{0^\circ}$" are higher than those for ``Noisy", indicating an improvement in the intelligibility and quality of noisy speech signals from the English dataset using the proposed approach.
Table \ref{tab:expI_tmhint} lists the STOI and PESQ scores of ``Noisy'', ``IANS$_{0^\circ}$'', ``STOI-NS$_{0^\circ}$'', and ``T-NSBF'' associated with the TMHINT database. From the table, the improved metric performances from ``Noisy" to ``IANS$_{0^\circ}$" confirm that the proposed IANS method can effectively enhance the intelligibility and sound quality of
noise-corrupted utterances. Notably, in both Tables \ref{tab:expI_wsj} and \ref{tab:expI_tmhint}, STOI-NS$_{0^\circ}$ has the highest STOI and PESQ scores on average, indicating that in these two experiments, if we properly choose the null angle to be $\phi_\mathrm{STOI-NS}^\star$, we generate results with even higher intelligibility and quality than the results from null-steering beamforming where we had the prior knowledge of the DOAs of the speech and interference signals.
One potential factor that may have influenced this outcome is the non-anechoic nature of the room, resulting in signals propagating through multiple pathways. Hence, nulling the angle $\theta_i$ may not be the optimal choice for STOI. 

\begin{table}[!b]\vspace{-0.6cm}
\caption{Average STOI and PESQ scores for ``Noisy'', ``IANS$_{0^\circ}$'', ``STOI-NS$_{0^\circ}$'', and ``T-NSBF'' on  WSJ}\label{tab:expI_wsj}
    \centering
    \begin{tabularx}{\columnwidth}{
    >{\centering}p{0.7cm}
    |>{\centering}p{1.1cm}
    |>{\centering}p{1.1cm}
    |>{\centering}p{1.7cm}
    |>{\centering\arraybackslash}p{1.8cm}
    }
    \toprule
    \hline
    & Noisy & IANS$_{0^\circ}$ & STOI-NS$_{0^\circ}$ & T-NSBF \\
    \hline
    
    STOI & 0.765 & 0.857 & \bf{0.862} & 0.858 \\
    PESQ & 1.277 & 1.538 & \bf{1.581} & 1.553  \\
    
    \hline
    \bottomrule
    \end{tabularx} \vspace{-0.4cm}
    
\end{table}
\begin{table}[!b]
\caption{Average STOI and PESQ scores for ``Noisy'', ``IANS$_{0^\circ}$'', ``STOI-NS$_{0^\circ}$'', and ``T-NSBF'' on TMHINT}\label{tab:expI_tmhint}
    \centering
    \begin{tabularx}{\columnwidth}{
    >{\centering}p{0.7cm}
    |>{\centering}p{1.1cm}
    |>{\centering}p{1.1cm}
    |>{\centering}p{1.7cm}
    |>{\centering\arraybackslash}p{1.8cm}
    }
    \toprule
    \hline
    & Noisy & IANS$_{0^\circ}$ & STOI-NS$_{0^\circ}$ & T-NSBF \\
    \hline
    
    STOI & 0.820 & 0.881 & \bf{0.895} & 0.892  \\
    PESQ & 1.417 & 1.614 & \bf{1.770} & 1.744 \\
    
    \hline
    \bottomrule    
    \end{tabularx} \vspace{-0.4cm}
\end{table}
\begin{table}[!b]
\caption{Average STOI and PESQ scores for ``IANS$_{0^\circ}$'', ``IANS$_{80^\circ}$'', ``STOI-NS$_{0^\circ}$'', and ``STOI-NS$_{80^\circ}$'' on WSJ }\label{tab:thetad_wsj}
    \centering
    \begin{tabularx}{\columnwidth}{
    >{\centering}p{0.7cm}
    |>{\centering}p{1.15cm}
    |>{\centering}p{1.15cm}
    |>{\centering}p{1.7cm}
    |>{\centering\arraybackslash}p{1.8cm}
    }
    \toprule
    \hline
    & IANS$_{0^\circ}$ & IANS$_{80^\circ}$ & STOI-NS$_{0^\circ}$ & STOI-NS$_{80^\circ}$ \\
    \hline
    
    STOI & 0.857 & 0.857 & 0.862 & 0.862 \\
    PESQ & 1.538 & 1.541 & 1.581 & 1.583 \\
    
    \hline
    \bottomrule
    \end{tabularx} \vspace{-0.4cm}
    
\end{table}
\begin{table}[!b]
\caption{Average STOI and PESQ scores for ``IANS$_{0^\circ}$'', ``IANS$_{80^\circ}$'', ``STOI-NS$_{0^\circ}$'', and ``STOI-NS$_{80^\circ}$'' on TMHINT}\label{tab:thetad_tmhint}
    \centering
    \begin{tabularx}{\columnwidth}{
    >{\centering}p{0.7cm}
    |>{\centering}p{1.15cm}
    |>{\centering}p{1.15cm}
    |>{\centering}p{1.7cm}
    |>{\centering\arraybackslash}p{1.8cm}
    }
    \toprule
    \hline
    & IANS$_{0^\circ}$ & IANS$_{80^\circ}$ & STOI-NS$_{0^\circ}$ & STOI-NS$_{80^\circ}$ \\
    \hline
   
    STOI & 0.881 & 0.881 & 0.895 & 0.895 \\
    PESQ & 1.614 & 1.615 & 1.770 & 1.771 \\
    
    \hline
    \bottomrule    
    \end{tabularx}
\end{table}

Next, we will further investigate how different values of $\psi$ affects the performance of ``IANS$_{\psi}$'' and ``STOI-NS$_{\psi}$.'' Specifically, we compared the results obtained using $\psi = 0^{\circ}$ and $\psi = 80^{\circ}$.
These results are presented in Tables \ref{tab:thetad_wsj} and \ref{tab:thetad_tmhint}, which correspond to the WSJ and TMHINT databases, respectively. 
From both tables, when comparing ``IANS$_{0^\circ}$" with ``IANS$_{80^\circ}$" and ``STOI-NS$_{0^\circ}$" with ``STOI-NS$_{80^\circ}$", we can see that, even though the results corresponding to $\psi = 80^{\circ}$ yield higher PESQ scores, there is essentially no difference in STOI.
This implies that the large $90^\circ$ difference between $\psi$ and $\theta_s$ has an insignificant effect on the ability of the IANS algorithm to generate intelligibility-enhanced results in our experiments.

\begin{figure}[!t]
    \centering
    \includegraphics[width=0.9\columnwidth]{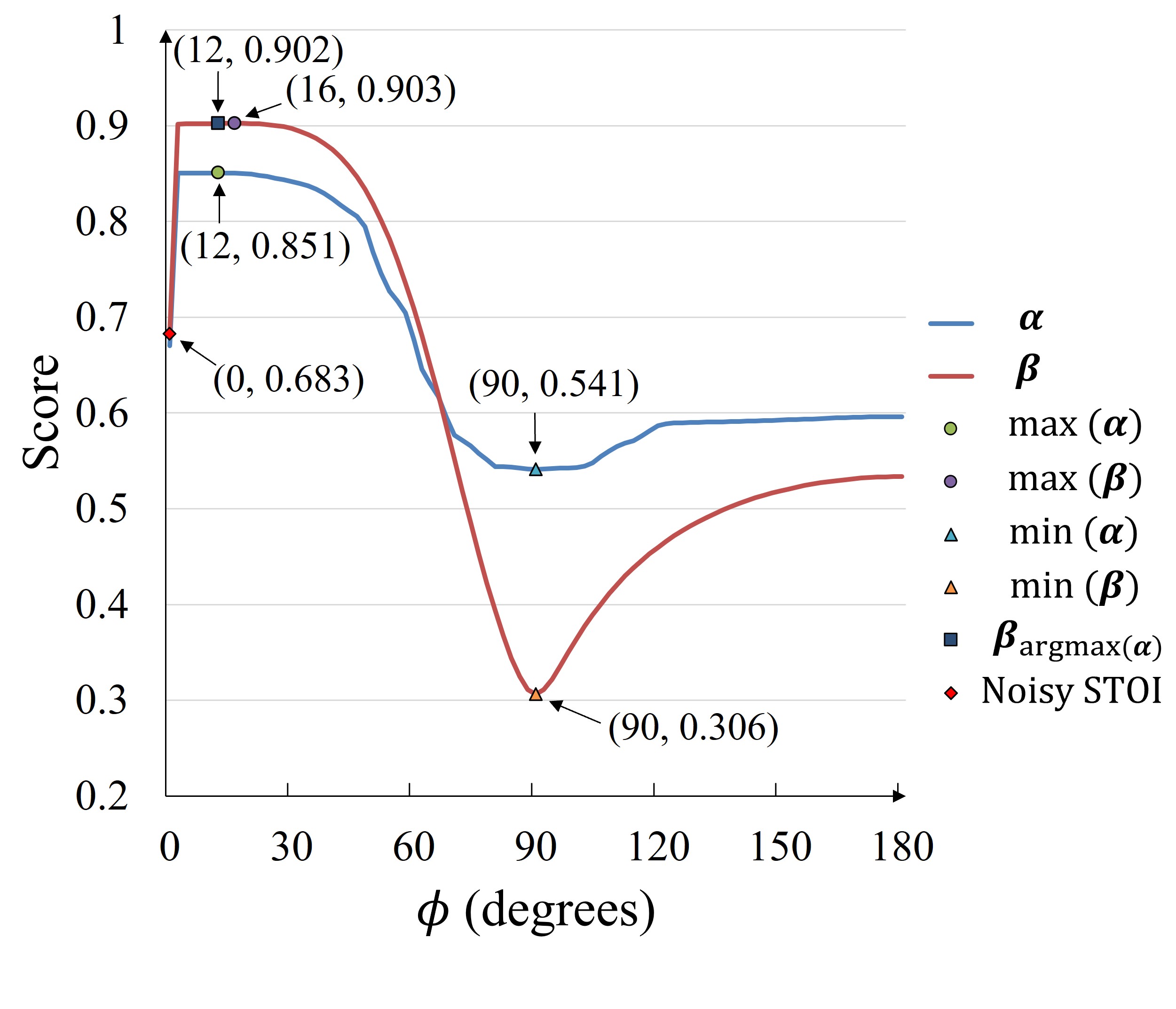}
     \vspace{-0.6cm}
    \caption{Comparing the STOI-Net score vector $\boldsymbol{\alpha}$ from IANS with STOI score vector $\boldsymbol{\beta}$ from STOI-NS in Scenario A using $\psi=0^{\circ}$} \vspace{-5mm}
    
    \label{fig:babble_PS}
\end{figure}

Finally, we present an additional analysis of $\boldsymbol{\alpha}$ and $\boldsymbol{\beta}$ in a particular scenario (Scenario A) to gain further insight into the similarities between the IANS and STOI-NS problems. The scenario consists of a female speaker from the WSJ dataset being interfered by the babble noise coming from a $22.5^{\circ}$ angle (i.e., $\theta_i=22.5^{\circ}$) with the SIR set to 0 dB. We let $\psi=0^{\circ}$ for both IANS and STOI-NS, which means that the STOI value in $\boldsymbol{\beta}$ corresponding to $\phi=0^{\circ}$ is the STOI score of the unprocessed signal at the reference microphone as we explained in Subsection \ref{subsec:NSBF}. The score values in both $\boldsymbol{\alpha}$ and $\boldsymbol{\beta}$ are represented by the two curves depicted in Fig. \ref{fig:babble_PS}.
The x-axis represents the values of $\phi$ in degrees, whereas the y-axis represents the values of $\boldsymbol{\alpha}$ and $\boldsymbol{\beta}$.
From the figure, we can see that the two curves have similar characteristics. Specifically, the lowest values for $\boldsymbol{\alpha}$ and $\boldsymbol{\beta}$ both correspond to $\phi=\theta_s=90^{\circ}$. Since both $\boldsymbol{\mathit{f}}_\mathrm{IANS}^\star$ and $\boldsymbol{\mathit{f}}_\mathrm{STOI-NS}^\star$ output results corresponding to the largest value in their respective score vectors, this suggests that they are both effective in preventing the speech signal from being severely attenuated in Scenario A.
Moreover, maximum values of $\boldsymbol{\alpha}$ and $\boldsymbol{\beta}$ occur at $\phi_\mathrm{IANS}^\star=12^\circ$ and $\phi_\mathrm{STOI-NS}^\star=16^\circ$, respectively. 
The corresponding STOI scores for 
$\boldsymbol{\mathit{f}}_\mathrm{IANS}^\star$ and $\boldsymbol{\mathit{f}}_\mathrm{STOI-NS}^\star$ are 0.902 (i.e., $\boldsymbol{\beta}_{\mathrm{argmax}(\boldsymbol{\alpha})}$) and 0.903 (i.e., $\mathrm{max}(\boldsymbol{\beta})$), respectively, which are at least 0.219 points higher than the STOI score of ``Noisy" at 0.683,  indicating the effectiveness in STOI enhancement of our IANS algorithm.

\section{Conclusion and future work}
\vspace{-0.5mm}
\label{sec:conclusion}
In this paper, we proposed a novel intelligibility-based optimization problem (i.e., the IANS problem) along with its corresponding enhancement system, the IANS beamformer. 
The system determines the optimal output speech with the highest intelligibility scores by combining the NSBF and STOI-Net modules, where NSBF processes the input recordings and STOI-Net provides STOI predictions.
We conducted experiments using cross-lingual datasets (Mandarin and English). The experimental results show that the proposed IANS system can effectively map the input signals to intelligibility and quality enhanced speech. It was also demonstrated that IANS produces robust performance regardless of whether the distortionless response is steered near the direction of the speech source.
In the future, we will evaluate the combination of beamforming systems with different evaluation modules, such as quality and mean opinion score assessment models \cite{zezario2022deep}, and test our system in more complex noisy environments. 
In addition, we will integrate single-channel speech enhancement methods with IANS to further enhance speech signals.

\bibliographystyle{IEEEbib}
\bibliography{mlsp}

\end{document}